\documentclass[twocolumn,showpacs,preprintnumbers,amsmath,amssymb]{revtex4}
\usepackage{graphicx}
\usepackage{dcolumn}
\usepackage{bm}

\begin{document}

\title{Stability of complex networks under the evolution of attack and repair}

\author{L.P. Chi}
\email{chilp@iopp.ccnu.edu.cn} \affiliation{Institute of Particle
Physics, Hua-Zhong (Central China) Normal University, Wuhan
430079, P.R. China}
\author{C.B. Yang}
\affiliation{Institute of Particle Physics, Hua-Zhong (Central
China) Normal University, Wuhan 430079, P.R. China}
\author{X. Cai}
\affiliation{Institute of Particle Physics, Hua-Zhong (Central
China) Normal University, Wuhan 430079, P.R. China}

\date{\today}

\begin{abstract}

With a simple attack and repair evolution model, we investigate
and  compare the stability of the Erd\"{o}s-R\'{e}nyi random
graphs (RG) and Barab\'{a}si-Albert scale-free (SF) networks. We
introduce a new quantity, \textit{invulnerability} $I(s)$, to
describe the stability of the system. We find that both RG and SF
networks can evolve to a stationary state. The stationary value
$I_{c}$ has a power-law dependence on the average degree $\langle
k \rangle_{rg}$ for RG networks; and an exponential relationship
with the repair probability $p_{sf}$ for SF networks. We also
discuss the topological changes of RG and SF networks between the
initial and stationary states. We observe that the networks in the
stationary state have smaller average degree $\langle k \rangle$
but larger clustering coefficient $C$ and stronger assortativity
$r$.

\end{abstract}


\maketitle

\section{Introduction}

Complex networks are an essential part of modern society. Many
social~\cite{wasserman}, biological~\cite{williams},
transportation~\cite{Li,Chi,Latora} and communication~\cite{BA1}
systems can be cast into the form of complex networks, a set of
nodes joined together by links indicating interactions. The
investigation of complex networks was initiated by Erd\H{o}s and
R\'{e}nyi~\cite{ER1} in 1950s. They introduced a kind of simplest
and most straightforward complex networks, described as random
graphs. In a random graph, $N$ nodes are connected randomly with
the probability $p_{rg}$. This model has guided our thinking about
complex networks for decades since its introduction. In 1999
Barab\'{a}si and Albert~\cite{BA2,BA3,BA4} began to put the
emphasis on the network dynamics and constructed networks with the
algorithm of \textit{growth} and \textit{preferential attachment}.
According to the BA model, the network grows over time by the
addition of new nodes and links. A node newly added to the network
randomly selects $m$ other nodes to establish new links, with a
selection probability that increases with the number of links of
the selected node. One of the most relevant is given by the
scale-free nature of the degree distribution $P(k)$, defined as
the probability that a randomly chosen node has degree $k$ (i.e.,
it is connected to other $k$ nodes). In mathematical terms, the
scale-free property translates into a power-law function of the
form $P(k) \sim k^{-\gamma}$.

Recently the security of complex networks to the random failures
or intentional attacks has attracted a great deal of
attention~\cite{Crucitti1,Crucitti2,LiXiang}. The random failure
is simulated as the deletion of network nodes or links chosen at
random, while intentional attack as the targeted removal of a
specific class of nodes or links. The work by Albert \textit{et
al}~\cite{attack1} demonstrated that scale-free networks, at
variance of random graphs, are robust against random failures of
nodes but vulnerable to intentional attacks. The analytical
approach~\cite{Cohen,Callaway} has also been developed for
investigating the error and attack tolerance.

However, despite the fine work of studies on network tolerance,
little effort has been made on the repair of the attacked
networks. In this paper, we propose a simple model and study the
evolution of complex networks under continuous attacks and
repairs. This effect can be pictured for the organism: the
organism has to experience various DNA damages and corresponding
repairs during its growth. The effect can be also exemplified by
the Internet: the development of Internet always follows the
contest of virus attacks and protections. Our model is aiming at
investigating the stability appearing in the evolution of complex
systems frequently attacked by internal or environmental agents.
The model shows that complex networks will become more stable in
the evolution and eventually reach the stationary states. We also
discuss the topological changes of complex networks before and
after the evolution. It will be useful for understanding the
dynamics of complex systems.

\section{Evolution model}

We start by constructing a network according to the
Erd\"{o}s-R\'{e}nyi random graph (RG) model or Barab\'{a}si-Albert
scale-free (SF) model. The algorithms of RG and SF networks have
been mentioned in the Introduction. The dynamics of our model is
defined in terms of the following two operations:

\begin{enumerate}

\item \textit{Attack}: Find a node with the maximum degree $k_{max}$ and remove all
its links. (If several nodes happen to have the same highest
degree of connection, we randomly choose one of them.)

\item \textit{Repair}: Reconnect this node
with the other nodes in the network with probability $p_{re}$.

\end{enumerate}

Then, the evolution comes into the next Monte Carlo time step. The
random repair is considered because most of the time our operation
to the system is more or less blinded due to incomplete
information. In addition, the size of the system remains unchanged
during the evolution since the way of simulated attacks is the
deletion of links. In order to reduce the number of parameters in
the model, we set the repair probability $p_{re}$ the same as the
connection probability $p_{rg}$ in RG networks. This set also
implies that the information does not increase from initial
construction to later repairing.

\section{Results and discussions}

\subsection{Stability of complex networks}

We have performed extensive numerical simulations to study the
evolution of RG and SF networks. It can be easily seen that the
maximum degree $k_{max}$ has a tendency to decrease in the
evolution. In Fig. 1, we give a snapshot of the maximum degree
$k_{max}$ versus time step $s$ with $N=$ 10 000 nodes and repair
probability $p_{re}=0.005$ for RG and SF networks. We find that
the fluctuation of $k_{max}$ is very large for RG networks. For SF
networks, $k_{max}$ decreases very steeply at first and then
steadily with $s$. The behavior of scale-free networks is rooted
in their extremely inhomogeneity in which the removal of a few
highly connected nodes dramatically reduces the connections of the
networks.

\begin{figure}
\begin{center}
\includegraphics[height=6cm,width=8cm]{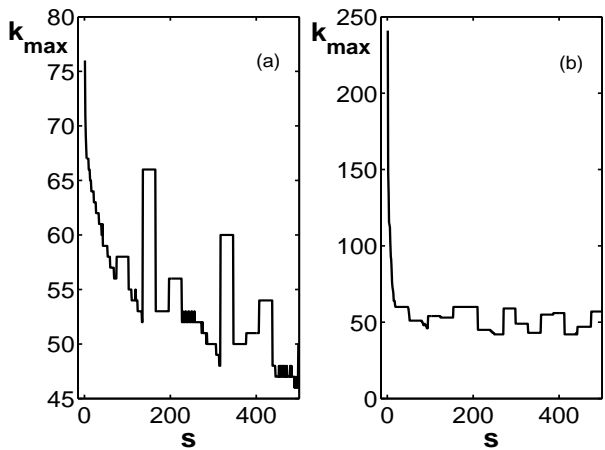}
\caption{Plots of $k_{max}$ versus time step $s$ with size $N=$ 10
000 and $p_{re}=0.005$ for (a) random graphs; (b) scale-free
networks.}
\end{center}
\label{fig1}
\end{figure}

From intuition, a node with less links to others will be attacked
less frequently. Thus a network with smaller maximum connection
degree is more stable. To describe the stability of the network,
we introduce a new quality, \textit{Invulnerability} $I(s)$. The
definition of $I(s)$ is analogous to that of the \textit{Gap}
$G(s)$ in the Bak-Sneppen evolution model~\cite{BS1,BS2}, which is
used to simulate the evolution of species. Considering an
evolution of network with maximum degrees $k_{max}(1)$,
$k_{max}(2)$,..., $k_{max}(s)$, invulnerability $I(s)$ at time $s$
is defined

\begin{equation}
I(s)=1/Min\{k_{\rm max}(i)\} \ \ \textbf{for\ \ \ } i\leq s\ ,
\end{equation}

\noindent i.e., the inverse of the minimum of all the maximum
degrees in the evolution before moment $s$. Initial value of
$I(s)$ is equal to $1/{k_{max}(1)}$. $I(s)$ reflects the attack
tolerance of the network. When $I(s)$ is small, the network is
vulnerable to attack and behaves less stability. Obviously from
definition, $I(s)$ is non-decreasing function of evolution, and
some fluctuations in $k_{max}$ have been filtered out.

\begin{figure}
\begin{center}
\includegraphics[width=8cm]{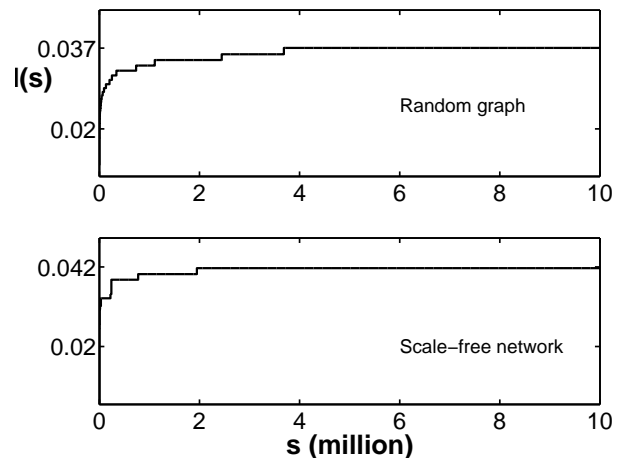}
\caption{Plots of $I(s)$ versus step $s (million)$ with $N=$ 10
000 and $p_{re}=0.005$.}
\end{center}
\label{fig2}
\end{figure}

In Fig. 2 we plot the evolution $I(s)$ versus $s (million)$ of RG
and SF networks under $N=$ 10 000 and $p_{re}=0.005$. We observe
that in both networks $I(s)$ increases very quickly at small $s$
but slowly at large $s$ and finally reaches a stationary value
$I_{c}$ when $s$ is large enough. The increase of $I(s)$ indicates
that the system is getting more and more stable in general by
experiencing continuous attacks and repairs. This figure is
similar to the envelope function of Bak-Sneppen evolution model.
Without interference from outside world, the network evolves to a
stationary state, and the process takes place over a very long
transient period.

\subsection{Stationary states of random graphs}

In this section we will present the properties of stationary value
$I_c^{rg}$ in the stationary state for RG networks. In RG
networks, the average degree $\langle k \rangle_{rg}$ is
determined by the connection probability $p_{rg}$ and the size
$N$, $\langle k \rangle_{rg} = p_{rg}N$. With $\langle k
\rangle_{rg} = 20$ fixed and $p_{re}=p_{rg}$, we find that
$I_{c}^{rg}$ stays unchanged at 0.2 with size $N$ ranging from 100
to 10 000. This result is interesting, because it shows that the
stationary value $I_{c}^{rg}$ depends not on the network size $N$,
but the average degree $\langle k \rangle_{rg}$.

\begin{figure}
\begin{center}
\includegraphics[width=6cm]{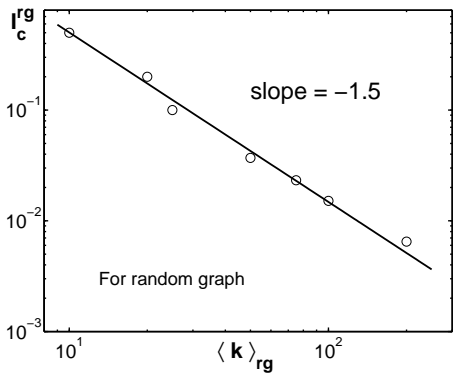}
\caption{Plot of $I_{c}^{rg}$ versus $\langle k \rangle_{rg}$ with
size $N$ ranging from 100 to 10 000.}
\end{center}
\label{fig3}
\end{figure}

To get the relationship between $I_{c}^{rg}$ and $\langle k
\rangle_{rg}$, we show in Fig. 3 $I_{c}^{rg}$ as a function of
$\langle k \rangle_{rg}$ in a log-log plot. We find that
$I_{c}^{rg}$ has a power-law dependence on the average degree
$\langle k \rangle_{rg}$,

\begin{equation}
I_{c}^{rg}(\langle k \rangle_{rg}) \ \propto \  \langle k
\rangle_{rg}^{-\tau},
\end{equation}

\noindent where the exponent $\tau$ is about 1.5. Fig. 3
illustrates that after the network has relaxed to the stationary
state, the stability of the network will increase rapidly with the
decrease of average degree $\langle k \rangle_{rg}$. Thus, when
the $\langle k \rangle_{rg}$ is small, i.e., less communications
and interactions in the network, the system will be more stable.

\subsection{Stationary states of scale-free networks}

The properties of stationary value $I_c^{sf}$ in the stationary
state for SF networks will be illustrated in this section. In Fig.
4, we plot $I_{c}^{sf}$ as a function of repair probability
$p_{sf}$ in a semi-log coordinate with $N=$ 1000 (circles); 5000
(squares) and 10 000 (triangles). We find that $I_c^{sf}$
decreases with the increase of $N$ or $p_{sf}$. While large $N$ or
$p_{sf}$ means the large number of links in networks. In other
words, the stability of the system relies on the internal
interactions within the system, that is, the stronger the
interactions, the less stable the system. From Fig. 4 we can also
observe that in this semi-log plot $I_{c}^{sf}$ has an
approximately linear decrease with $p_{sf}$ under fixed $N$, i.e.,
an exponential relationship between $I_c^{sf}$ and $p_{sf}$,

\begin{equation}
I_{c}^{sf}(p_{sf}) \ \sim \ \exp({-a(N)p_{sf}+b(N)}),
\end{equation}

\noindent where $a(N)$ and $b(N)$ are two fitting parameters
related to the size $N$.

\begin{figure}
\begin{center}
\includegraphics[width=6cm]{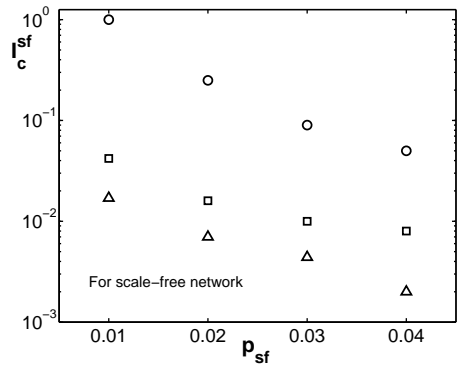}
\caption{Plot of $I_{c}^{sf}$ versus repair probability $p_{sf}$
with size $N=$ 1000 (circles); 5000 (squares); and 10 000
(triangles).}
\end{center}
\label{fig4}
\end{figure}

\subsection{Topological changes of complex networks}

In order to offer further information of complex networks in the
stationary states, we compare some structural properties of the
networks in the initial states with those in the stationary
states. In Table~\ref{comparison}, we display the topological
changes of RG and SF networks under $N=$ 10 000 and
$p_{re}=0.005$. We use the superscript $i$ and $s$ to distinguish
the initial and stationary states.

\begin{table}
\caption{\em Comparison of topological changes in initial and
stationary states between RG and SF networks under $N=$ 10 000 and
$p_{re}=0.005$.}
\begin{center}
\begin{tabular}{c|c|cc|cc|cc|cc} \hline\hline

     &$I_{c}$ &$\langle k \rangle^i$&$\langle k \rangle^s$&$L^i$&$L^s$&$C^i$&$C^s$&$r^i$&$r^s$ \\
\hline
RG   & 0.037 &50.16 &19.34 &2.77 &3.68 &0.005  &0.57  &0.00041 &0.35 \\
\hline
SF   & 0.042 &4.0   &3.35  &5.09 &7.17 &0.0039 &0.014 &-0.049  &0.39 \\
\hline

\end{tabular}
\end{center}
\label{comparison}
\end{table}

\textit{Average degree}. We find that the average degree in
stationary states $\langle k \rangle ^ s$ decreases dramatically
for RG networks but slightly for SF networks. This result is
rooted in the homogeneity of RG networks. Since in RG networks all
nodes have approximately the same number of links,  majority of
nodes in the network will be impacted by the attacks and, hence,
the rapid decrease of $\langle k \rangle$.

\textit{Characteristic path length}. The characteristic path
length $L$ is the average length of the shortest paths between any
two nodes in the network. $L^s$ in the stationary state is a
little larger than $L^i $, which exhibits that it will take a
little longer for the nodes in the network to communicate with
each other after the evolution.

\textit{Clustering coefficient}. The clustering coefficient of
node $i$ is defined as the existing numbers $n_{i}$ among the
links of node $i$ over all the possible links, that is,
$C_{i}=n_{i}/[k_{i}(k_{i}-1)/2]$. The clustering coefficient of
the whole network is $C=\frac{1}{N} \sum C_{i}$. $C^s$ for RG and
SF networks are increased by $99\%$ and $72\%$, compared with
$C^i$ in the initial states. It suggests that the nodes are highly
clustered in the stationary states.

\textit{Assortativity}. The assortativity $r$ in the range $-1
\leq r \leq 1$ is another interesting feature of complex networks.
A network is said to show assortative, $r>0$, if the high-degree
nodes in the network preferentially connect to other high-degree
nodes. By contrast, a network is disassortative, $r<0$, if the
high-degree nodes tend to connect to other low-degree nodes. The
values of $r^i$ and $r^s$ have shown that both RG and SF networks
turn to strongly assortative networks in the stationary states
from initial weakly assortative or even disassortative ones.

\textit{Degree distribution}. Fig. 4(a) and 4(b) display the
degree distributions $P(k)$ of RG and SF networks in the initial
and stationary states under $N=$ 10 000 and $p_{re}=0.005$. For RG
networks, the degree distribution $P(k)_{rg}^s$ in the stationary
state is a Gaussian one with mean value $\langle k \rangle_{rg}^s
\simeq 19$ and standard deviation $\sigma_{rg}^s \simeq 7$, while
the $P(k)_{rg}^i$ in the initial state is a Poisson distribution
with $\langle k \rangle_{rg}^i \simeq 50$ and
$\sigma_{rg}^i=\sqrt{\langle k \rangle_{rg}^i} \simeq \sqrt{50}$.
For SF networks, $P(k)_{sf}^i$ and $P(k)_{sf}^s$ are both
power-laws, $P(k)_{sf} \sim k_{sf}^ {-\gamma}$, with the similar
exponent about 2.4. The differences between $P(k)_{sf}^i$ and
$P(k)_{sf}^s$ are that the fat tail in $P(k)_{sf}^i$ vanishes; and
the maximum degree decreases from the initial value of 241 to 50
in stationary state. Besides, Fig. 4(a) and 4(b) show that the
shape of the degree distributions is almost the same for RG and SF
networks in initial and stationary states. It suggests that the
general structures of the systems do not change even after a long
time evolution to the stationary states.

\begin{figure}
\begin{center}
\includegraphics[height=6cm,width=8cm]{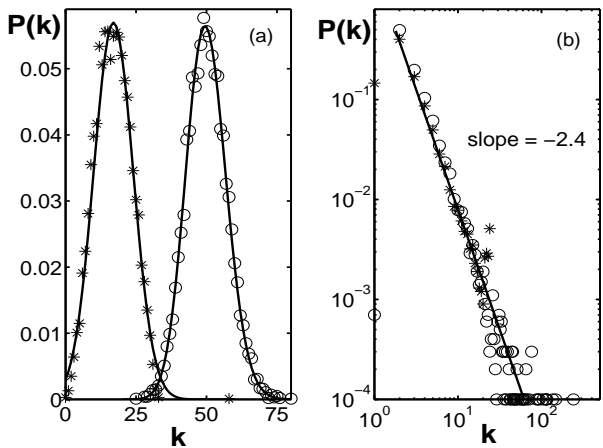}
\caption{A comparison of degree distributions under $N=$ 10 000
and $p_{re}=0.005$ in initial (circles) and stationary (asterisks)
states for (a) random graphs; and (b) scale-free networks . Solid
lines are the fit curves. }
\end{center}
\label{fig5}
\end{figure}

In general, the system in the stationary state has smaller average
degree $\langle k \rangle$ but larger clustering coefficient $ C$
and stronger assortativity $r$. It implies that the system after
undergoing a long period of evolution is composed of small highly
clustered clusterings, which well fits the natural world. In
ecosystems, agents autonomously form communities, in which
individuals are highly correlated. Commercial airlines, for
reasons of efficiency, prefer to have a small number of hubs where
all routes would connect.

\section{Conclusions}

In this paper we study the stability of random graphs and
scale-free networks with a simple attack and repair evolution
model. We introduce a new quality, \textit{invulnerability}
$I(s)$, to describe the stability of the networks. A stationary
state with fixed $I(s)$ is observed during the evolution of the
network. The stationary value of invulnerability $I_{c}$ is found
to have a power-law dependence on $\langle k \rangle_{rg}$ for RG
networks; and an exponential relationship with repair probability
$p_{sf}$ for SF networks. We give further information on the
evolution of the networks by comparing the topological changes
between the initial and stationary states. We conclude that the
networks in the stationary states are composed of small highly
clustered clusterings, which well fits the evolution of complex
systems in natural world.

Still, there are many issues to be addressed, such as the
correlations and the fluctuations in $k_{max}$ in the evolution,
especially after the stationary state. These fluctuations may tell
us more about the nature of the stationary state. The behavior of
the average degree in the evolution also may shed some light on
the dynamics of the networks. In addition, it is worthwhile to
investigate whether the evolution of networks demonstrate
self-organized criticality (SOC), according to the similarity
between the evolution of invulnerability $I(s)$ with that of the
envelope function $G(s)$ in Bak-Sneppen model. All these topics
can not be covered in this paper, and will be discussed later.

\section*{Acknowledgments}

This work was supported in part by the National Natural Science
Foundation of China under grant No. 70271067 and by the Ministry
of Education of China under grant No. 03113.


\begin{thebibliography}{99}
\bibitem{wasserman}
S. Wasserman and K. Faust, \textit{Social Network Analysis},
(Cambridge University Press, Cambridge, England, 1994).

\bibitem{williams}
R. J. Williams and N. D. Martinez, Nature \textbf{404} (2000) 180.

\bibitem{Li}
W. Li and X. Cai, Phys. Rev. E \textbf{69} (2004) 046106.

\bibitem{Chi}
L. P. Chi, R. Wang, H. Su, \textit{et al}, Chin. Phys. Lett.
\textbf{20} (2003) 1393.

\bibitem{Latora}
V. Latora and M. Marchiori, Physica A \textbf{314} (2002) 109.

\bibitem{BA1}
R. Albert, H. Jeong and A.-L. Barab\'{a}si, Nature \textbf{401}
(1999) 130.

\bibitem{ER1}
P. Erd\H{o}s and A. R\'{e}nyi, Publ. Math. Inst. Hung. Acad. Sci.
\textbf{5} (1960) 17.

\bibitem{BA2}
A.-L. Barab\'{a}si and R. Albert, Science \textbf{286} (1999) 509.

\bibitem{BA3}
A.-L. Barab\'{a}si, R. Albert, and H. Jeong, Physica A
\textbf{281} (2000) 69.

\bibitem{BA4}
R. Albert and A.-L. Barab\'{a}si, Rev. Mod. Phys. \textbf{74}
(2002) 47.

\bibitem{Crucitti1}
P. Crucitti, V. Latora, M. Marchiori and A. Rapisarda, Physica A
\textbf{340} (2004) 388.

\bibitem{Crucitti2}
P. Crucitti, V. Latora, M. Marchiori and A. Rapisarda, Physica A
\textbf{320} (2003) 622.

\bibitem{LiXiang}
X. Li and G. Chen, Physica A \textbf{328} (2003) 274.

\bibitem{attack1}
R. Albert, H. Jeong and A.-L. Barab\'{a}si, Nature \textbf{406}
(2000) 378.

\bibitem{Cohen}
R. Cohen, K. Erez, D. ben-Avraham and S. Havlin, Phys. Rev. Lett.
\textbf{86} (2000) 4626.

\bibitem{Callaway}
D.S. Callaway, M.E.J. Newman, S.H. Strogatz and D.J. Watts, Phys.
Rev. Lett. \textbf{85} (2000) 5468.

\bibitem{BS1}
P. Bak and K. Sneppen, Phys. Rev. Lett. \textbf{71} (1993) 4083.

\bibitem{BS2}
M. Paczuski, S. Maslov and P. Bak, Phys. Rev. E \textbf{53} (1996)
414.

\end{thebibliography}
\end{document}